\documentclass[5p, times]{elsarticle}
\makeatletter
\def\ps@pprintTitle{%
   \let\@oddhead\@empty
   \let\@evenhead\@empty
   \let\@oddfoot\@empty
   \let\@evenfoot\@oddfoot
}
\makeatother

\usepackage{bm}
\usepackage[svgnames,dvipsnames,table]{xcolor}
\usepackage[colorlinks]{hyperref}
\usepackage{amsmath}
\usepackage{amssymb}
\usepackage{adjustbox}
\usepackage{multirow}
\usepackage{tikz}
\usepackage{pgfplots}
\usepgfplotslibrary{patchplots}
\usetikzlibrary{trees,shapes,snakes,calc,matrix,positioning,fit,arrows,patterns,backgrounds,quotes,arrows.meta}
\usepackage{siunitx}
\usepackage{multicol}
\usepackage{caption}
\usepackage{subcaption} 
\usetikzlibrary{shapes.geometric}
\usetikzlibrary{shapes.arrows}
\usepgfplotslibrary{fillbetween}
\usepackage{array}
\usepackage{tabularx}
\usepackage{lipsum} 
 \usepackage{collcell}
\usepackage{hhline}
\usetikzlibrary{spy, calc}
\usepackage{array}
\usepackage{tabularx}
\usepackage{tabulary}
\usepackage{xspace}

\usepackage{bbding}
\usepackage{pifont}
\usepackage{diagbox}
\usepackage{booktabs}
\usepackage{makecell}
\usetikzlibrary{fit}
\usepackage{rotating}
\usepgfplotslibrary{colorbrewer}
\usepackage{fontawesome}
\usepackage{lscape}
\usepackage[para,online,flushleft]{threeparttable}
\usepackage{longtable}
\usepackage{tablefootnote}
\usepackage{makecell}
\usepackage{breakcites}
\usepackage{pifont}
\usepackage{hyphenat}
\begin{document}
\begin{frontmatter}

\title{Generating Synthetic Contrast-Enhanced Chest CT Images from Non-Contrast Scans Using Slice-Consistent Brownian Bridge Diffusion Network}


\author[1]{Pouya Shiri}
\ead{pouya.shiri@usask.ca}
\author[1]{Xin Yi}
\ead{xin.yi@usask.ca}
\author[1]{Neel P. Mistry}
\ead{neel.mistry@usask.ca}
\author[3]{Samaneh Javadinia}
\ead{samanehjavadinia@uvic.ca}
\author[3]{Mohammad Chegini}
\ead{mchegini@uvic.ca}
\author[2]{Seok-Bum Ko}
\ead{seokbum.ko@usask.ca}

\author[3]{Amirali Baniasadi}
\ead{amiralib@uvic.ca}

\author[1]{Scott J. Adams}
\ead{scott.adams@usask.ca}

\address[1]{Department of Medical Imaging, University of Saskatchewan, 103 Hospital Dr, Saskatoon, SK, S7N 0W8 Canada}
\address[2]{Department of Electrical and Computer Engineering, University of Saskatchewan, 57 Campus Dr, Saskatoon, SK, S7N 5A9 Canada}
\address[3]{Department of Electrical and Computer Engineering, University of Victoria, 3800 Finnerty Rd, Victoria, BC, V8P 5J2 Canada}

\begin{abstract}
Contrast-enhanced computed tomography (CT) imaging is essential for diagnosing and monitoring thoracic diseases, including aortic pathologies. However, contrast agents pose risks such as nephrotoxicity and allergic-like reactions. The ability to generate high-fidelity synthetic contrast-enhanced CT angiography (CTA) images without contrast administration would be transformative, enhancing patient safety and accessibility while reducing healthcare costs. In this study, we propose the first bridge diffusion-based solution for synthesizing contrast-enhanced CTA images from non-contrast CT scans. Our approach builds on the Slice-Consistent Brownian Bridge Diffusion Model (SC-BBDM), leveraging its ability to model complex mappings while maintaining consistency across slices. Unlike conventional slice-wise synthesis methods, our framework preserves full 3D anatomical integrity while operating in a high-resolution 2D fashion, allowing seamless volumetric interpretation under a low memory budget. To ensure robust spatial alignment, we implement a comprehensive preprocessing pipeline that includes resampling, registration using the Symmetric Normalization method, and a sophisticated dilated segmentation mask to extract the aorta and surrounding structures. We create two datasets from the Coltea-Lung dataset: one containing only the aorta and another including both the aorta and heart, enabling a detailed analysis of anatomical context. We compare our approach against baseline methods on both datasets, demonstrating its effectiveness in preserving vascular structures while enhancing contrast fidelity.
\end{abstract}


\end{frontmatter}

\section{Introduction}

Computed tomography angiography (CTA) has become a crucial tool for the diagnosis and monitoring of acute aortic syndromes, chronic aortic dissection, and aortic aneurysms, as well as pre-procedural planning and follow-up for aortic stent-grafts. Accurate diagnosis and monitoring of aortic disease requires the administration of iodine-based contrast agents, which pose risks such as nephrotoxicity and allergic-like reactions. These concerns limit its widespread applicability, particularly for patients with renal impairment or hypersensitivity to contrast media. The ability to synthesize contrast-enhanced images from non-contrast CT scans presents a transformative opportunity to mitigate these risks while improving accessibility and reducing healthcare costs.

Diffusion models have demonstrated remarkable success in various medical imaging tasks, including cross-modality synthesis \cite{xie2021,kim2021} and low-dose CT enhancement \cite{Zhang2023,Zhou2022}. Among these, Latent Diffusion Models (LDMs) \cite{rombach2022high} have been particularly effective in compressing data into latent spaces, enabling computationally efficient generation. However, when applied to 3D volumetric data, LDMs still encounter significant memory constraints, often requiring over 24GB of VRAM for high-resolution synthesis. These limitations motivate our focus on 2D-based solutions that can be extended to 3D volumetric representations without excessive computational overhead.

Recent advancements in Brownian Bridge Diffusion Models (BBDMs) and bridge diffusion frameworks offer a promising alternative to standard Denoising Diffusion Probabilistic Models (DDPMs) \cite{ho2020denoising}. Unlike DDPMs, which rely on forward and reverse Markov processes, BBDMs explicitly constrain the sampling trajectory between two endpoints using Brownian bridge mechanisms. This approach provides enhanced control over the diffusion process, leading to improved spatial consistency and anatomical fidelity, which are essential characteristics for medical imaging applications. The Slice-Consistent BBDM (SC-BBDM) \cite{choo2024} extends BBDMs with mechanisms such as Style Key Conditioning (SKC) and Inter-Slice Trajectory Alignment (ISTA) to enhance medical image synthesis. SKC ensures uniformity in imaging styles across slices, while ISTA promotes spatial coherence, reducing slice-wise inconsistencies. By integrating these techniques with controlled noise sampling, SC-BBDM effectively generates high-quality synthetic contrast\hyp{}enhanced chest CT volumes.

Existing methods for contrast-enhanced CT synthesis include generative adversarial networks (GANs) and multi-task frameworks. Xiong et al. propose a cascaded multi-task generative framework for detecting aortic dissection from Non-Contrast (NCT) scans \cite{xiong2022cascaded}. Their model segments the aorta, synthesizes CTA, and classifies aortic dissection cases, achieving high sensitivity (0.938). However, their dataset is private, comparisons are restricted to Pix2Pix, and their reported metrics include zero voxels, which creates a discrepancy between visual and numerical performance. Similarly, Yin et al. \cite{yin2024} introduced a Multi-stage Cascade Generative Adversarial Network (MCGAN) designed for synthesizing CTA images of the aorta. Their approach focused on improving the synthesis of intimal flaps, which are critical for diagnosing aortic dissection. While their model incorporates dense residual attention blocks (DRAB) for shallow feature extraction and a UNet-based architecture for deep feature extraction, the reliance on adversarial training makes the model susceptible to mode collapse and instability, limiting its generalizability to broader clinical applications.

\textbf{Contributions}. Our key contributions are as follows: We develop a robust preprocessing pipeline with image resampling and registration to align non-contrast and contrast-enhanced scans. A dilated segmentation mask is implemented to focus synthesis on vascular structures. We construct two datasets from the Coltea-Lung dataset \cite{cytran2022} i.e. one containing only the aorta and another including both the aorta and heart, to enable detailed anatomical analysis. Our approach is comprehensively compared with baseline synthesis models, evaluating anatomical fidelity, slice consistency, and clinical applicability. To ensure high-quality training data, a clinical specialist manually reviewed and filtered suboptimal cases from the Coltea-Lung dataset. Finally, we introduce the first application of bridge diffusion models for contrast enhancement in CT imaging, offering a safer alternative to iodine-based contrast agents.




\section{Method}
\subsection{BBDM}
BBDM extends conventional diffusion models by enabling controlled transitions between arbitrary source and target distributions, rather than being limited to Gaussian noise. This makes it well-suited for image-to-image (I2I) translation. We define the following notations: $\mathbf{X}, \mathbf{Y}, \mathbf{X_t} \in \mathbf{R}^{Z \times H \times W}$ represent the complete CTA, NCT, and latent volumes, respectively. The individual slices at index $i$ of these volumes are denoted as $x^i, y^i, x^i_t \in \mathbf{R}^{H \times W}$. Additionally, we define sub-volumes centered around the $i^{th}$ slice, including its $2N$ adjacent slices, as $\mathbf{X}^i, \mathbf{Y}^i, \mathbf{X}^i_t \in \mathbf{R}^{(2N+1) \times H \times W}$. 
\\

\textbf{Forward and Reverse Process}: 
Given a source-target image pair \( (\mathbf{x}, \mathbf{y}) \), the forward diffusion process follows:
\begin{equation}
    q_{BB}(\mathbf{x}_t | \mathbf{x}_0, \mathbf{y}) = \mathcal{N}(\mathbf{x}_t; (1 - m_t) \mathbf{x}_0 + m_t \mathbf{y}, \delta_t \mathbf{I}),
\end{equation}
where \( m_t = t/T \) interpolates between the source \( \mathbf{x}_0 \) and the target \( \mathbf{y} \), while \( \delta_t = 2s(m_t - m_t^2) \) controls the variance, and $s > 0$ is a variance scaling hyperparameter that determines the noise magnitude in the Brownian bridge process. The reverse process, modeled by a neural network, follows:
\begin{equation}
    p_{\theta} (\mathbf{x}_{t-1} | \mathbf{x}_t, \mathbf{y}) = \mathcal{N}(\mathbf{x}_{t-1}; \mu_{\theta} (\mathbf{x}_t, t), \tilde{\delta}_t \mathbf{I}),
\end{equation}
where \( \mu_{\theta} (\mathbf{x}_t, t) \) is the predicted mean, and $\tilde{\delta}_t$ is the reverse process variance, typically set equal or proportional to $\delta_t$ for simplicity or numerical stability.\\

\textbf{Training Objective}: 
The model is trained to align the learned reverse process with the true forward process by minimizing:
\begin{equation}
    \mathbf{E}_{\mathbf{x}_0, \mathbf{y}, \boldsymbol{\epsilon}} \left[ \| c_{\epsilon t} m_t (\mathbf{y} - \mathbf{x}_0) + \sqrt{\delta_t} \boldsymbol{\epsilon} - \boldsymbol{\epsilon}_{\theta} (\mathbf{x}_t, t) \|^2_2 \right],
\end{equation}
where \( \boldsymbol{\epsilon} \sim \mathcal{N}(0, \mathbf{I}) \) is sampled Gaussian noise, and \( \boldsymbol{\epsilon}_{\theta} \) is the noise prediction model.

\subsection{SC-BBDM}
\begin{figure*}[htp]
    \centering
    \includegraphics[width=0.8\textwidth]{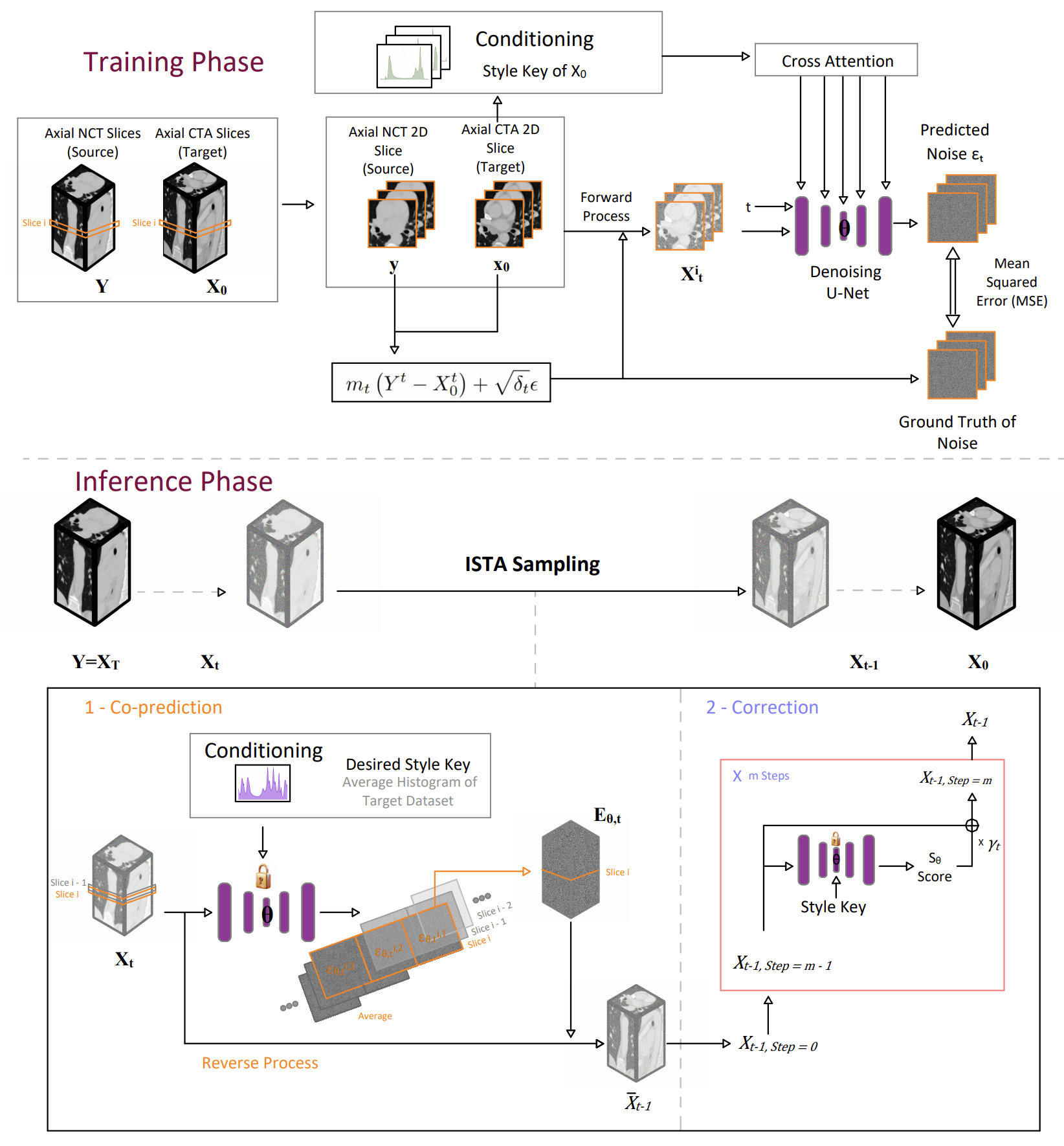}
    

    
    \caption{Overview of SC-BBDM: (Top) Training phase, where the model learns from data, and (Bottom) Inference phase, where the trained model is applied to new inputs.}
    \label{fig:sc-bbdm-overview}
\end{figure*}

SC-BBDM builds on top of BBDM by introducing two key concepts: Style Key Conditioning (SKC) and Inter-Slice Trajectory Alignment (ISTA). Figure \ref{fig:sc-bbdm-overview} shows the architecture of SC-BBDM for training and inference.

\textbf{SKC}: While training BBDM slice-by-slice, inconsistencies in brightness and contrast may emerge due to variations in the target images. To address this, SKC is introduced. A style key is derived from the intensity histogram of a target volume, serving as an invariant descriptor of its appearance. This ensures consistency across generated images while allowing controlled synthesis. The diffusion objective is modified as:
\begin{equation}
\mathbf{E}_{\mathbf{X}_0^i, \mathbf{Y}^i, \boldsymbol{\epsilon}} \left[ \| c_{\epsilon t} m_t (\mathbf{Y}^i - \mathbf{X}_0^i) + \sqrt{\delta_t} \boldsymbol{\epsilon} - \boldsymbol{\epsilon}_{\theta} (\mathbf{X}_t^i, \mathbf{S}_{\text{style}}, t) \|^2_2 \right]
\end{equation}
where \( \mathbf{S}_{\text{style}} \) represents a set of histograms capturing the target volume's intensity distribution, and \( c_{\epsilon t} \) is a time-dependent coefficient derived from the Brownian Bridge posterior, scaling the guiding term \( m_t (\mathbf{Y}^i - \mathbf{X}_0^i) \) relative to the noise component.\\

\textbf{ISTA}: SKC ensures global consistency but does not fully address local inconsistencies between adjacent slices. To enforce inter-slice coherence, ISTA is introduced. It improves consistency by aggregating predictions from adjacent slices. Given an input slice \( \mathbf{X}_t \), overlapping predictions are combined to form a refined estimate:
\begin{equation}
\mathbf{E}_{\theta, t} = \text{CP}_\theta (\mathbf{X}_t, \mathbf{S}_{\text{style}}, t) = \left[ \bar{\epsilon}_{\theta, t}^1, \bar{\epsilon}_{\theta, t}^2, \dots, \bar{\epsilon}_{\theta, t}^Z \right],
\end{equation}
where \( \bar{\epsilon}_{\theta, t}^i \) represents the predicted noise estimate for the \( i \)th slice. To align predictions with the data manifold, a correction step is applied:
\begin{equation}
\mathbf{X}_{t, m} = \mathbf{X}_{t, m-1} + \lambda \delta_{t|t-1} (\sqrt{d} / \|\mathbf{S}_{\theta}\|_2)^2 \mathbf{S}_{\theta}
\end{equation}
where \( \mathbf{X}_{t,0} = \bar{\mathbf{X}}_{t-1} \) is the initial estimate; \( \lambda \) is a step-size hyperparameter; \( d \) controls the correction magnitude, and \( \mathbf{S}_{\theta} \) enforces trajectory alignment.

\subsection{Dataset}




The Coltea-Lung CT dataset~\cite{cytran2022} consists of 100 anonymized triphasic lung CT scans collected from female patients, including a total of 37290 images. The dataset documentation does not indicate which pathologies, if any, are present in the images, and our radiologist’s visual review did not reveal any obvious pathological findings. For each patient, three CT scans are provided, corresponding to the non-contrast (native), arterial, and venous phases of the same anatomical region. This triphasic representation offers a unique opportunity to evaluate imaging techniques designed for cross-phase image generation.


\subsection{Pre-processing}

\begin{figure*}[tbp]
    \centering
    \resizebox{\textwidth}{!}{
    \includegraphics[width=\linewidth]{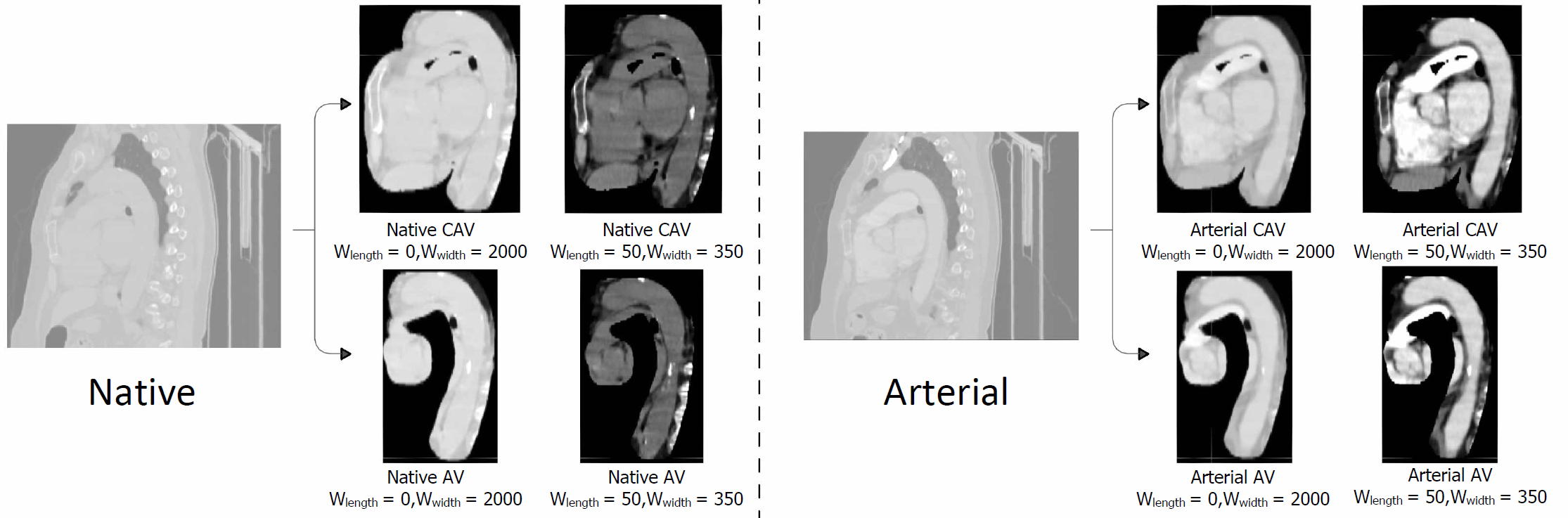}}
    \caption{Pre-processing for native and arterial CT images. The original inputs are processed into two datasets: CAV and AV, where cropping is performed using dilated segmentation masks. The sample is also visualized under two windowing settings: Full HU range and Soft Tissue Window, highlighting different tissue structures.}
    \label{fig:coltea-sample}
\end{figure*}

In this work, we use the SyN (Symmetric Normalization) registration method \cite{AVANTS2008} to align the non-contrast and contrast-enhanced images. Before registration, we removed structures unrelated to the aorta and its immediate cardiac context such as the spine and abdominal organs using segmentation-based cropping. This step ensured that registration focused on the thoracic vascular region and reduced interference from irrelevant anatomy.
After registration, to provide the model with contextual information about tissue characteristics, we retained controlled surrounding regions by applying mask dilation rather than restricting the input to the aorta alone. Figure~\ref{fig:coltea-sample} presents the sagittal view of the middle slice of one sample along with the pre-processed results. The original input scans are processed into two datasets, Cardiac-Aortic Volume (CAV) and Aortic Volume (AV), using segmentation masks to simulate different level of background tissue around the aorta. The AV dataset includes only the aorta and its surrounding region, defined by a 10 voxel dilation of the aorta segmentation mask obtained via the TotalSegmentator tool \cite{totalseg}. The CAV dataset extends this region to include the myocardium, left and right atria, and left and right ventricles, with a 20 voxel distance dilation. For both native and arterial images, we apply the union of all arterial and native segmentation masks to ensure consistency across datasets.

We removed a total of 38 cases from the Coltea-Lung dataset which did not meet the standards for diagnostic accuracy. In particular, these cases contained significant noise and artifact that would hinder diagnostic interpretation, ultimately affecting clinical management. Afterwards, we did a 80/10/10 split on the rest of the data. We resampled the images to 256×256 in the axial plane with a uniform voxel spacing of [0.7, 0.7, 2.5] mm. We clipped the intensity values to a HU range of -1000 to 1000 and normalized to [0,1] for consistency in deep learning models.

\subsection{Implementation Details}
We use a 3090 GPU with 24GB of memory to train the SC-BBDM. The model was trained on axial slices for 100 epochs, completing in approximately 3 days with the diffusion model undergoing 1000 diffusion steps. We adopt a linear schedule for $m_t$, where it increases uniformly from 0.001 to 0.999 accross timesteps. 

\begin{table}[!ht]
\centering
\caption{UNet configuration for denoising.}
\label{tab:unet_config}
\begin{tabular}{ll}
\hline
\textbf{Parameter} & \textbf{Value} \\
\hline
Image size          & $256 \times 256$ \\
In/Out channels & $6$/$3$ \\
Base channels       & $128$ \\
Channel multipliers & $(1, 4, 8)$ \\
Residual blocks     & $2$ per level \\
Attention resolutions & $\{32, 16, 8\}$ \\
Attention heads     & $8$ ($64$ channels each) \\
\hline
\end{tabular}
\end{table}

The denoising network is based on OpenAI's UNet architecture, configured as summarized in Table~\ref{tab:unet_config}.

For inference, we employed 50 DDIM \cite{ddim} ISTA sampling steps with a correction factor of $M=1$. The style key, represented by histograms, is derived from the ground-truth equivalent (arterial phase) during training and computed as the averaged histogram of all training samples during inference. Inference takes about 23 minutes per case.

We observed that inference results were not always consistent with validation loss across checkpoints. To address this, we integrated full-volume validation sampling into the training process and monitored NRMSE, PSNR, and SSIM to select the best model.

\subsection{Baselines}
We chose Pix2Pix \cite{pix2pix} and CyTran \cite{cytran2022} as baselines for our study. CyTran was selected because it initially presented the Coltea-Lung dataset and achieved strong performance on contrast-enhanced data. Pix2Pix was chosen for its effectiveness in image-to-image translation tasks, making it a solid baseline. Both methods provide valuable points of comparison for evaluating our approach.

\subsection{Evaluation Methods}
\textbf{Quantitative results}:

We evaluate performance using three commonly used image similarity metrics: Peak Signal-to-Noise Ratio (PSNR), Structural Similarity Index Measure (SSIM), and Normalized Root Mean Squared Error (NRMSE). All metrics are computed over entire 3D volumes. Additionally, metrics are reported on non-zero voxels of both the ground truth and prediction, indicated as (NZ) in Table~\ref{tab:performance_comparison}, to ensure the evaluation focuses on clinically relevant regions. Avg-Train denotes inference using the average style keys from training, while GT-Test uses the ground-truth style key for each test sample.

As shown in Table~\ref{tab:performance_comparison}, the AV dataset consistently achieves better scores than CAV, for example PSNR improves from around 32.5 dB on CAV to 36.8 dB on AV for SC-BBDM GT-Test, reflecting the lower complexity of AV. The difference between Avg-Train and GT-Test configurations of SC-BBDM is small, such as 32.461 vs. 32.578 PSNR on CAV and 36.245 vs. 36.820 PSNR on AV, yet visual inspection confirms that GT-Test offers higher perceptual quality.

When comparing SC-BBDM and CyTran, the differences vary by metric and dataset. On CAV, both methods report close PSNR (32.578 vs. 32.730) and same NRMSE (0.024), but CyTran achieves higher SSIM (0.953 vs. 0.937). On AV, SC-BBDM outperforms CyTran in PSNR (36.820 vs. 33.807) and NRMSE (0.015 vs. 0.022), while CyTran retains an advantage in SSIM (0.975 vs. 0.932). For non-zero voxels, SC-BBDM shows a clear edge: on AV its PSNR is 27.480 compared to CyTran’s 23.690, and its NRMSE is 0.044 compared to 0.069, indicating stronger accuracy in clinically relevant region, and on CAV it is lagging behind CyTran by a small margin (27.694 vs. 27.867 for PSNR, and 0.834 vs. 0.851 for SSIM).

%
\begin{table*}[tbp]
    \centering
    \setlength{\tabcolsep}{6pt} 
    \renewcommand{\arraystretch}{1.2} 
    \resizebox{\textwidth}{!}{%
    \begin{tabular}{llccc|ccc}
        \hline
        \textbf{Dataset} & \textbf{Method} & \textbf{NRMSE} $\downarrow$ & \textbf{PSNR} $\uparrow$ & \textbf{SSIM}$\uparrow$ & \textbf{NRMSE (NZ)} $\downarrow$ & \textbf{PSNR (NZ)} $\uparrow$ & \textbf{SSIM (NZ)} $\uparrow$ \\
        \hline
        \multirow{4}{*}{\textbf{CAV}}   
        & SC-BBDM (Avg-Train) & 0.025 & 32.461 & 0.936 & 0.045 & 27.579 & 0.837 \\
        & SC-BBDM (GT-Test) & \textbf{0.024} & 32.578 & 0.937 & \textbf{0.043} & 27.694 & 0.834 \\
        & CyTran \cite{cytran2022} & 0.024 & \textbf{32.730} & \textbf{0.953} & 0.043 & \textbf{27.867} & \textbf{0.851} \\
        & Pix2Pix \cite{pix2pix} & 0.032 & 30.319 & 0.933 & 0.057 & 25.454 & 0.829 \\
        \hline
        \multirow{4}{*}{\textbf{AV}}  
        & SC-BBDM (Avg-Train) & 0.016 & 36.245 & 0.926 & 0.048 & 26.771 & \textbf{0.846} \\
        & SC-BBDM (GT-Test) & \textbf{0.015} & \textbf{36.820} & 0.932 & \textbf{0.044} & \textbf{27.480} & 0.843 \\
        & CyTran \cite{cytran2022} & 0.022 & 33.807 & 0.975 & 0.069 & 23.690 & 0.786 \\
        & Pix2Pix \cite{pix2pix} & 0.021 & 33.889 & \textbf{0.980} & 0.069 & 23.727 & 0.836 \\
        \hline
    \end{tabular}%
    }
    \caption{Performance comparison of different methods on CAV and AV datasets. Avg-train uses the average histograms of training samples, while GT-test derives the style key from the ground-truth histograms of test samples.}
    \label{tab:performance_comparison}
\end{table*}

\textbf{Qualitative results}:
To gain deeper insight, generated images were reviewed by a radiologist alongside the ground truth. To ensure consistency, we applied appropriate windowing to the ground-truth CTA, input NCT, and generated CTA with a window width of 350 HU and a window level of 50 HU. This adjustment highlights the relevant anatomic structures, allowing for a more accurate assessment of contrast enhancement, particularly in the aorta and heart regions. We have provided videos showing slice\hyp{}wise performance of SC\hyp{}BBDM compared to the ground\hyp{}truth, along with the baselines (CyTran and Pix2Pix), and the input NCT volumes in the Supplementary Material of this paper.

Figure~\ref{fig:side_by_side} is divided into AV slices (left) and CAV views (right) and groups SC-BBDM's predictions into three bands: Good, i.e. virtually indistinguishable from the contrast ground truth; moderate, i.e. minor shape or intensity deviations that still leave anatomy recognizable; and bad, i.e. pronounced artifacts that obscure or distort vascular structure. In AV moderate case shows exaggerated fat planes between the main pulmonary trunk and aortic arch (top) and artifactual narrowing of the descending aorta (bottom). Bad cases exhibit blurring of the anatomical borders (as outlined) due to a combination of artifact and hyperenhancement. In CAV,  moderate case demonstrates blurring of anatomical borders and loss of cardiac chambers due to hyperenhancement, and bad cases exhibit artifactual narrowing of the aorta and its branches, as well as blurring of cardiac chamber walls.  Despite these view‑specific shortcomings, SC-BBDM outputs preserve vessel geometry and intensity distributions more faithfully than CyTran and Pix2Pix, whose images are marred by grid‑pattern texture and streak artefacts.

\begin{figure*}[btp]
    \centering
    \begin{minipage}{0.48\textwidth}
        \centering
        \includegraphics[width=\textwidth]{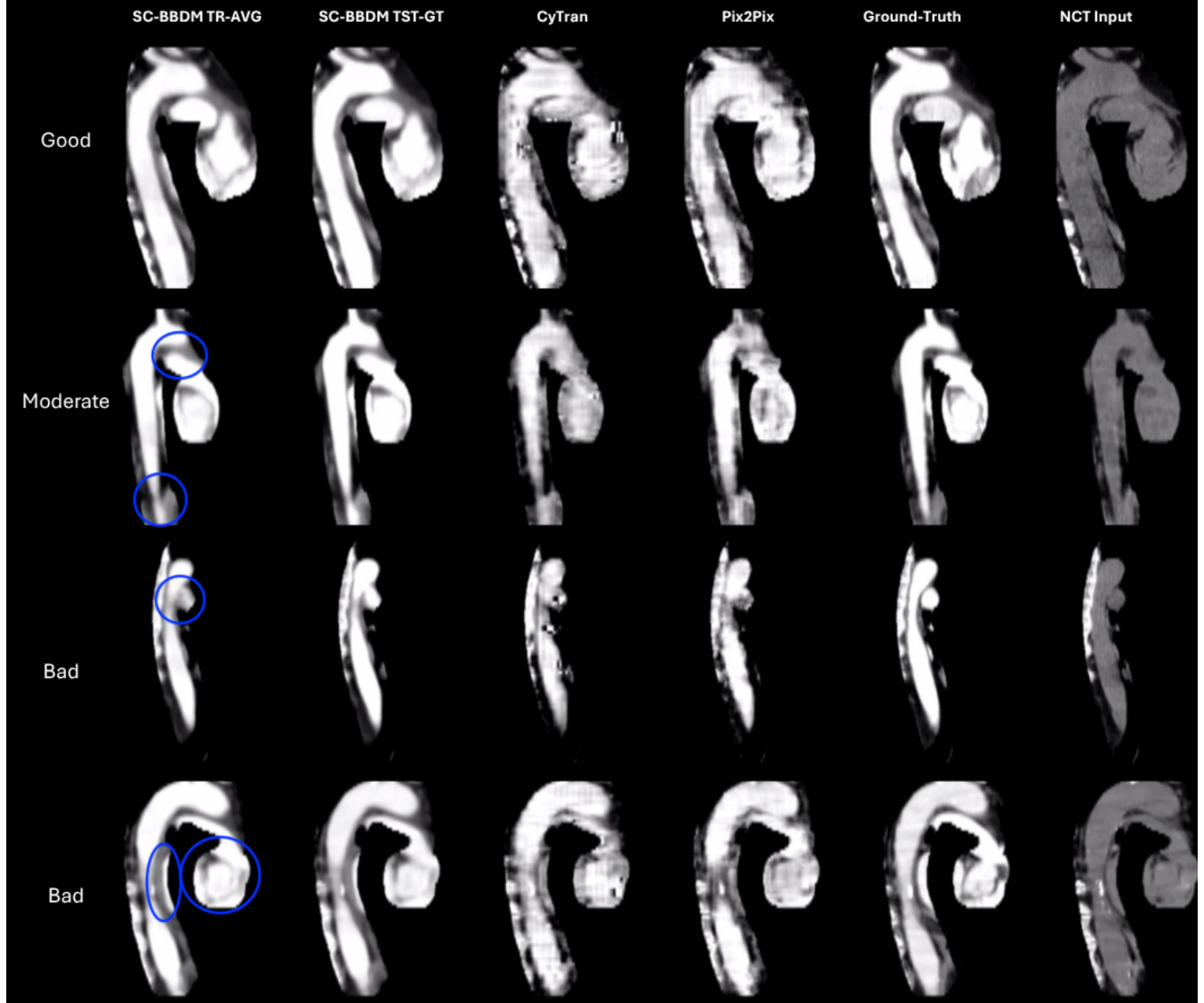}
        \label{fig:sm_res_vis}
    \end{minipage}
    \hfill
    \begin{minipage}{0.48\textwidth}
        \centering
        \includegraphics[width=\textwidth]{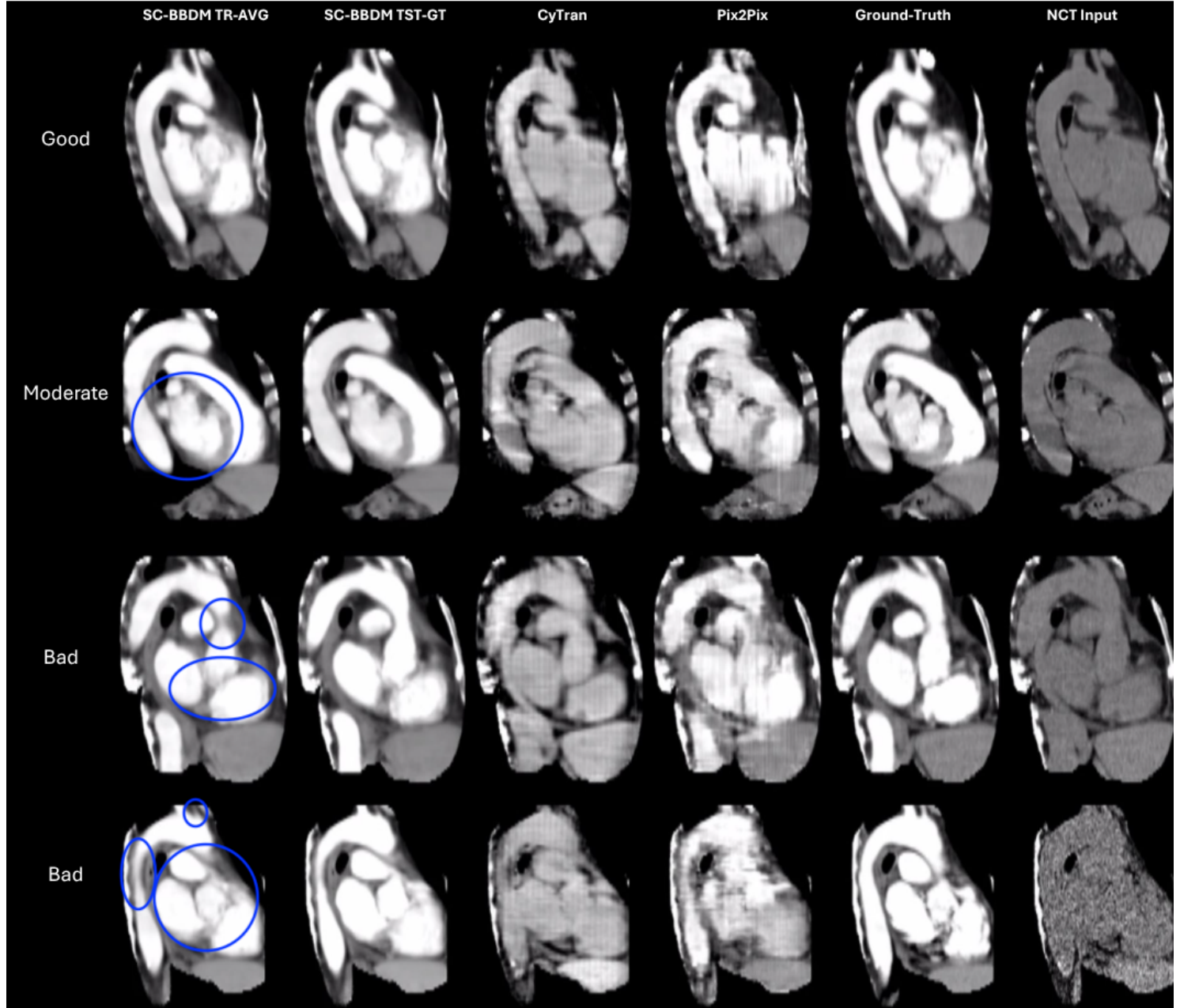}
        \label{fig:big_res_vis}
    \end{minipage}
    \caption{Comparison of synthesized images using different methods: the AV dataset on the left side, and the CAV dataset on the right side. Columns represent different approaches. TR-AVG means using the average style key of training, and TST-GT means using the ground-truth style key of test samples. Blue circles highlight notable artifacts or discrepancies.}
    \label{fig:side_by_side}
\end{figure*}

A promising future direction for evaluating visual quality is the integration of clinically meaningful metrics alongside traditional image similarity scores. Assessments should incorporate measurements routinely used in practice, such as aortic cross-sectional areas, luminal diameters at standardized anatomical landmarks, true and false lumen dimensions in dissection cases, and aneurysm volumes. These metrics provide a direct link between image synthesis quality and clinical decision-making. Ultimately, moving toward task-specific evaluation where the performance of downstream clinical measurements and risk stratification tools on synthetic data mirrors that on ground truth, will ensure that visual improvements translate into clinically actionable accuracy.

To address the perceptual quality gap between TR-AVG and TST-GT, one strategy is to increase diversity in the training set through 3D data augmentation techniques that simulate a wider range of contrast appearances, reducing the reliance on precise style cues. Another approach is to incorporate perceptual or adversarial losses that emphasize structural and visual fidelity, making the network less sensitive to style variation.

\section{Conclusion}
This study proposes a bridge diffusion-based solution for generating synthetic contrast-enhanced chest CT images from non-contrast scans. By leveraging SC-BBDM’s slice-consistency mechanisms, our approach preserves anatomical structures while maintaining spatial coherence across slices. A robust preprocessing pipeline, including rigid registration and segmentation-based masking, further improved alignment and synthesis quality. Experimental results on the filtered Coltea-Lung dataset demonstrated that SC-BBDM outperforms baseline methods such as CyTran and Pix2Pix, particularly in preserving vascular structures and contrast fidelity. While numerical metrics provided a general assessment, expert review highlighted the importance of qualitative evaluation under clinically relevant windowing settings.

Future work will focus on improving model robustness across diverse patient populations and conducting clinical validation on aortic pathologies. By providing a safer alternative to contrast-enhanced imaging, our approach has the potential to enhance diagnostic accessibility while reducing healthcare costs.

\bibliographystyle{splncs04}


\begin{thebibliography}{10}
\providecommand{\url}[1]{\texttt{#1}}
\providecommand{\urlprefix}{URL }
\providecommand{\doi}[1]{https://doi.org/#1}

\bibitem{AVANTS2008}
Avants, B., Epstein, C., Grossman, M., Gee, J.: Symmetric diffeomorphic image registration with cross-correlation: Evaluating automated labeling of elderly and neurodegenerative brain. Medical Image Analysis  \textbf{12}(1),  26–41 (Feb 2008). \doi{10.1016/j.media.2007.06.004}, \url{http://dx.doi.org/10.1016/j.media.2007.06.004}

\bibitem{choo2024}
Choo, K., Jun, Y., Yun, M., Hwang, S.J.: Slice-consistent 3d volumetric brain ct-to-mri translation with 2d brownian bridge diffusion model. arXiv preprint arXiv:2301.08815  (2024)

\bibitem{cytran2022}
Hassan, T., Arnaud, F., Liang, S.: Cytran: Cycle-consistent transformers for non-contrast to contrast ct translation. arXiv preprint arXiv:2110.06400  (2022), dataset details available at \url{https://arxiv.org/abs/2110.06400}

\bibitem{ho2020denoising}
Ho, J., Jain, A., Abbeel, P.: Denoising diffusion probabilistic models. Advances in Neural Information Processing Systems  \textbf{33},  6840--6851 (2020)

\bibitem{pix2pix}
Isola, P., Zhu, J.Y., Zhou, T., Efros, A.A.: Image-to-image translation with conditional adversarial networks (2016). \doi{10.48550/ARXIV.1611.07004}, \url{https://arxiv.org/abs/1611.07004}

\bibitem{kim2021}
Kim, S.W., Kim, J.H., Kwak, S., Seo, M., Ryoo, C., Shin, C.I., Jang, S., Cho, J., Kim, Y.H., Jeon, K.: The feasibility of deep learning-based synthetic contrast-enhanced ct from nonenhanced ct in emergency department patients with acute abdominal pain. Scientific Reports  \textbf{11}(1) (Oct 2021). \doi{10.1038/s41598-021-99896-4}, \url{http://dx.doi.org/10.1038/s41598-021-99896-4}

\bibitem{rombach2022high}
Rombach, R., Blattmann, A., Lorenz, D., Esser, P., Ommer, B.: High-resolution image synthesis with latent diffusion models. Proceedings of the IEEE/CVF Conference on Computer Vision and Pattern Recognition pp. 10684--10695 (2022)

\bibitem{ddim}
Song, J., Meng, C., Ermon, S.: Denoising diffusion implicit models (2020). \doi{10.48550/ARXIV.2010.02502}, \url{https://arxiv.org/abs/2010.02502}

\bibitem{totalseg}
Wasserthal, J., Breit, H.C., Meyer, M.T., Pradella, M., Hinck, D., Sauter, A.W., Heye, T., Boll, D.T., Cyriac, J., Yang, S., Bach, M., Segeroth, M.: Totalsegmentator: Robust segmentation of 104 anatomic structures in ct images. Radiology: Artificial Intelligence  \textbf{5}(5) (Sep 2023). \doi{10.1148/ryai.230024}, \url{http://dx.doi.org/10.1148/ryai.230024}

\bibitem{xie2021}
Xie, H., Lei, Y., Wang, T., Patel, P., Curran, W.J., Liu, T., Tang, X., Yang, X.: Generation of contrast-enhanced ct with residual cycle-consistent generative adversarial network (res-cyclegan). In: Bosmans, H., Zhao, W., Yu, L. (eds.) Medical Imaging 2021: Physics of Medical Imaging. p.~141. SPIE (Feb 2021). \doi{10.1117/12.2581056}, \url{http://dx.doi.org/10.1117/12.2581056}

\bibitem{xiong2022cascaded}
Xiong, X., Ding, Y., Sun, C., Zhang, Z., Guan, X., Zhang, T., Chen, H., Liu, H., Cheng, Z., Zhao, L., Ma, X., Xie, G.: A cascaded multi-task generative framework for detecting aortic dissection on 3-d non-contrast-enhanced computed tomography. IEEE Journal of Biomedical and Health Informatics  \textbf{26}(10),  5177--5188 (2022). \doi{10.1109/JBHI.2022.3190293}

\bibitem{yin2024}
Yin, J., Peng, J., Li, X., Ju, J., Wang, J., Tu, H.: Multi-stage cascade gan for synthesis of contrast enhancement ct aorta images from non-contrast ct. Scientific Reports  \textbf{14}(1) (Oct 2024). \doi{10.1038/s41598-024-73515-4}, \url{http://dx.doi.org/10.1038/s41598-024-73515-4}

\bibitem{Zhang2023}
Zhang, W., Zhou, Z., Gao, Z., Yang, G., Xu, L., Wu, W., Zhang, H.: Multiple adversarial learning based angiography reconstruction for ultra-low-dose contrast medium ct. IEEE Journal of Biomedical and Health Informatics  \textbf{27}(1),  409–420 (Jan 2023). \doi{10.1109/jbhi.2022.3213595}, \url{http://dx.doi.org/10.1109/JBHI.2022.3213595}

\bibitem{Zhou2022}
Zhou, Z., Gao, Y., Zhang, W., Bo, K., Zhang, N., Wang, H., Wang, R., Du, Z., Firmin, D., Yang, G., Zhang, H., Xu, L.: Artificial intelligence–based full aortic ct angiography imaging with ultra-low-dose contrast medium: a preliminary study. European Radiology  \textbf{33}(1),  678–689 (Jul 2022). \doi{10.1007/s00330-022-08975-1}, \url{http://dx.doi.org/10.1007/s00330-022-08975-1}

\end{thebibliography}

\end{document}